\documentclass[]{article}
\usepackage{authblk}

\def\bea{\begin{eqnarray}}
\def\eea{\end{eqnarray}}

\def\pp{\mbox{$p$-$p$}}

\def\auau{\mbox{Au-Au}}

\def\pbpb{\mbox{Pb-Pb}}
\def\ppb{\mbox{$p$-Pb}}

\def\aa{\mbox{A-A}}
\def\nn{\mbox{N-N}}

\def\k0s{$\text{K}^0_\text{S}$}

\title{Undiscovering Quark-gluon Plasma}
\author{Thomas A.\ Trainor}
\affil{University of Washington, Seattle, WA 98195}

\date{\today}

\begin{document}

\maketitle

\begin{abstract}

Quark-gluon plasma (QGP) is a conjectured state of matter wherein the quarks and gluons of QCD theory, usually confined within the nucleons in atomic nuclei, are deconfined under conditions of extremely high temperature and pressure. 
According to the current understanding of big-bang evolution, conditions compatible with QGP formation existed for a brief early period as the universe expanded and cooled. In that scenario  present matter in the universe in some sense emerged from ``freezeout'' of the primordial QGP. This article describes the rise {and fall} of a community effort to create QGP in the laboratory and study its properties.
\end{abstract}

\section{QCD history}

Quantum chromodynamics (QCD) is the currently-accepted theory of the nuclear force or strong interaction that binds atomic nuclei together and determines their properties. QCD is similar in its structure to quantum electrodynamics (QED), the modern theory of electromagnetic interactions among electric charges and currents. A brief history of QED may establish context.

The theory of electromagnetism relied on pioneers like Benjamin Franklin to identify the existence and properties of electric charge (c.\ 1750) and of Michael Faraday and Andr\'e-Marie Amp\`ere to establish the relations among charges and currents (c.\ 1840). James Maxwell combined those experimental results with a ``field theory'' intuited by Faraday to formulate ``Maxwell's Equations'' (c.\ 1860) that would represent classical electromagnetism for the next forty years.

With the turn of the  twentieth century came special relativity by Albert Einstein (1905) and quantum mechanics emerging over two decades from a number of contributors (Bohr, Schroedinger, Heisenberg, Born). Effective coupling of relativity to quantum mechanics first took place with Paul Dirac's eponymous equation (1928) that described the coupling of charges to photons propagating the electromagnetic field and predicted the existence of {\em antimatter}. By the late forties a complete {\em relativistic quantum field theory} of electromagnetism, QED, had been formulated by Richard Feynman, Julian Schwinger and Shinichiro Tomonaga. QED included a description of the creation and annihilation of particle-antiparticle pairs. So-called Feynman diagrams were a central component of necessary pair bookkeeping that remains essential up to the present. 

A theory for the strong interaction relied in part on technical development of increasingly energetic particle accelerators -- cyclotrons, synchrotrons and particle colliders. With increasing collision energy came discovery of a growing array of strongly-interacting particles or {\em hadrons,} suggesting the presence of new underlying degrees of freedom as in the case of the chemical elements. The ``charges'' of the new theory, {\em quarks}, were discovered theoretically by organizing hadrons into groups (eight-fold way), by Murray Gell-Mann and George Zweig (1964), and experimentally via electron scattering off protons (1968, Stanford). 

QCD, a theory for the strong interaction analogous to QED but based on quarks and gluons, emerged in several stages during the seventies. 
Gluons are for QCD the equivalent of photons for QED, but unlike photons gluons carry (color) charge and interact with other gluons as well as quarks. The ``chromo'' in quantum chromodynamics refers to a requirement within QCD of three charge states, not two as for QED. QCD charge states are then  represented metaphorically by the RGB color system, with quarks in hadrons (baryons at least) required to sum to R+G+B = white. QCD is  a central element of the current Standard Model that describes the strong, electromagnetic and weak (but not gravitational) interactions. The quarks and gluons of QCD are  conjectured to coexist within a QGP under extreme conditions of temperature and pressure.

\section{Bevalac pioneers}

During the nineteen seventies interest in the study of heavy ion nuclear collisions increased substantially. An existing heavy ion linear accelerator (HILAC) was rebuilt as the SuperHILAC (c.\ 1972) to accelerate ions up to uranium nuclei. The SuperHILAC was in turn adapted (c.\ 1974) as an injector into the Bevatron (in more-recent usage BeV is replaced by GeV), a synchrotron constructed (c.\ 1954) specifically to confirm Dirac's prediction of antiparticles for the case of protons (antiprotons were  successfully detected in 1955). The conjoined Bevalac was capable of accelerating ions up to an energy of 2 GeV per nucleon. As such, collisions could be described as ``relativistic'' since accelerated nucleon kinetic energies were comparable to their rest-mass energies of about 1 GeV.

Successful achievement of relativistic nuclear collisions presented new challenges. Each nucleus-nucleus collision could be the source of many particles distributed over a wide range of angles. Conventional low-energy nuclear physics detector technology assumed detection of only a few particles. To accommodate the abundance of particle production newer technologies had to be developed, including a plastic ball (scintillator) detector~\cite{plasticball} and a streamer chamber~\cite{streamer}. The latter served as a replacement for massive bubble chambers based on liquid hydrogen. The former was based on a scintillator (light-generating)  array capable of identifying low-mass charged particles (protons, deuterons, alphas).

Production of large numbers of particles in each nucleus-nucleus (\mbox{A-B}) collision  challenged not only measurement technology but also statistical analysis methods. Angular distributions of several particle species required species identification and reduction to mathematical representations that retained most of the information carried by data {\em in an interpretable form}. An example is the {\em sphericity tensor}~\cite{sphericity}, an array of numbers derived from particle momenta and describing an ovoid (egg) shape representing the intensity of emitted particles distributed on two angles. Such analysis revealed {\em collective flow} of nucleons apparently in response to pressures developed during nucleus-nucleus collisions~\cite{bevflow}. 

The discovery of nucleon flow as reported in 1984 was a signal achievement of the Bevalac program. The possibility to determine the thermodynamic properties of a dense system of nucleons within heavy-ion collisions was also of great interest to astrophysicists studying the properties of {\em neutron stars}, remnants of supernova explosions wherein an extreme gravitational field forces atomic electrons into nuclei and what results is mostly neutrons. It is possible that the cores of neutron stars may be composed of QGP as a limiting case. The combined interests led to a proposal for construction of a {\em collider accelerator} dedicated to heavy-ion collisions and an attempt to produce QGP in collisions.

\section{Training days}

The proposed relativistic heavy-ion collider installation~\cite{ludlum1984}, dubbed RHIC (pronounced ``rick''), would require massive new resources in terms of funding, detector construction and the manpower required to construct and manage detectors, implement novel analysis methods and interpret their results, {\em and} guide the transformation of scientific results to published papers. Since this was a new field of study the required manpower would come mainly from university cyclotron and van de Graaff laboratories previously engaged mainly in low-energy nuclear physics. The new community would thus require extensive training by ``experts'' mainly associated with government laboratories such as Lawrence Berkeley National laboratory (LBNL) and Brookhaven National laboratory (BNL). 

The scope of the project is reflected in the detector complement ultimately installed: two large and two small detectors,  initially specializing in disjoint aspects of heavy-ion physics as it was then imagined. One of the large detectors installed, the STAR detector, weighed 1200 tons. Its cylindrical solenoid magnet was about 20 ft in diameter and 25 feet long. The largest active detector element was a time projection chamber (TPC) about 13 feet in diameter and 14 feet long. The collaboration size was more than 500 persons from about fifteen countries. The total manpower for the four RHIC collaborations was about 1500 scientists not counting support staff within the various institutions.

In retrospect, two issues built into that system posed risks to effective scientific research. As the main sources of information for the ``training'' period, a number of  noted experts had a common origin relating to the Bevalac heavy ion program and shared a scientific context derived from ``relativistic'' nuclear collisions wherein almost all nucleons remain intact as the principal degrees of freedom. In contrast, during the eighties and nineties high-energy physics (HEP) made great progress in understanding {\em nucleon} structure and hadronic interactions at collider energies 100-1000 times the Bevalac equivalent, where individual ``partons'' (quarks and gluons) are the dominant degrees of freedom.

A second risk related again to a few experts with a common background influencing a much larger number of inexperienced scientists as to statistical analysis methods and {\em interpretation of results}. What phenomena emerging from data, represented graphically in what way, might reveal the presence of a QGP? In effect, certain prejudices based on the Bevalac experience might be amplified by such asymmetry and given undeserved weight within later scientific debate.

\section{Early RHIC results}

The RHIC experimental program began in late summer of 2000. The first RHIC run, at reduced collision energy 130 GeV (200 GeV was the design goal), was the trial run of an extremely complex system operating for the first time. Despite a number of small breakdowns in collider and detectors a substantial amount of useful data emerged and led to significant published results. That achievement may be considered a technical and social triumph for the new RHIC community.

During the training period methods required to identify a QGP were discussed; what ``signals'' obtained from data might identify QGP formation? Discussions were  framed in the context of the Bevalac experience and  thermodynamics of a population of hadrons emerging from a conjectured QGP. Statistical analysis was mainly based on elementary concepts available from standard texts. 

An alternative approach was followed by what became the {\em Event Structure} working group within the STAR collaboration. A survey of state-of-art statistical methods including information theory and topology was pursued as a foundation. No {\em a priori} assumptions (no ``signals'') were imposed on data analysis and interpretation. Incorporation of recent HEP results where relevant was considered a priority. In particular, manifestations of {\em jets}, closely-correlated groups of hadrons {\em radiated} from scattered partons, became a central issue. A two-component (soft + hard) model (TCM) of pt spectrum structure was inferred from event multiplicity dependence of spectrum shapes~\cite{ppprd}. The two components were later associated with basic hadron production mechanisms.

\section{RHIC white papers}

RHIC results within the first three years of operation were summarized in collaboration ``white papers'' during 2004. To briefly summarize here the STAR white paper~\cite{starwhite} and 200 GeV \auau\ results, it is noted that hadron species abundances follow a prediction from a so-called statistical model assuming a thermalized particle source. However, no example is given for a collision system that does not follow such a trend. Hadron spectra on transverse momentum pt are fitted over a {\em limited momentum range} with a preferred spectrum model assuming production from a thermalized particle source (QGP) and properties of a conjectured medium are inferred, but available {\em alternative} analyses of such data spectra are not considered. Measurement of ``elliptic flow'' via parameter v2 is consistently given great emphasis. An indicator is the statement ``[t]he measured yields with respect to the reaction plane [v2] are among the most important results to date from RHIC: they provide critical hints of the properties of the bulk matter [QGP] at early stages. They indicate that it [QGP] behaves collectively, and is consistent with rapid (i.e., very short mean free path) attainment of at least approximate local thermal equilibrium in a QGP phase.''

In contrast, certain Event Structure measurements of jet-related angular correlation without imposing any selection criteria, {as reported in the STAR white paper}, revealed that low-energy jets survive even in central \auau\ collisions although their angular structure is altered. That result was in direct conflict with  formation of dense ``bulk matter'' QGP within \auau\ collisions. Differential spectrum analysis leading to a two-component spectrum model revealed that some preferred model fits assuming a thermalized emission system were actually describing substantial contributions from jet production. Inferred thermodynamic and hydrodynamic parameter values appeared to be misleading.

Certain observations from the STAR white paper (2005) are of interest.

``In our judgment...it is premature to conclude definitively that the matter produced in central RHIC collisions is a Quark-Gluon Plasma, as this term has been
understood by the scientific community for the past 20 years.'' 
...
``The indirect evidence for a thermodynamic transition and for attainment of local thermal equilibrium in the matter produced at RHIC are [sic] intertwined in the hydrodynamics [theory] account for observed hadron spectra and elliptic flow results.''
...
``The theory remains a patchwork of different treatments applied in succession to each stage of the collision evolution, without yet a clear delineation of the different aspects as distinct limits of one overarching, seamless theory.''
...
``The bottom line is that in the absence of a direct  `smoking gun' signal of deconfinement {\em revealed by experiment alone} [emphasis added], a QGP discovery claim must rest on the comparison [of data] with a promising, but still not yet mature, theoretical framework.''

The chaotic nature of the theory-vs-data comparisons in 2004 is apparent upon careful reading of that text (consistent with most of the white paper). Available data were at that point fragmentary and poorly understood, and what posed as theory (configured {\em ab initio} to accommodate a QGP scenario) was a confection of {\em nonpredictive ad hoc components} tuned to match selected data. Nevertheless, broad enthusiasm for discovery of  QGP became the {\em vox populi}.

Comments on plans for the future included ``RHIC measurements should concentrate on verifying and extending its new observations of jet quenching and v2 scaling...'' thus identifying issues considered most important for QGP claims. Some specifics: ``[e]stablish v2 scaling more definitively'' and ``[e]stablish that jet quenching is an indicator of parton, and not hadron, energy loss.'' Event Structure methods have indeed been applied privately to those two questions over the past twenty years with very interesting results, as described below.

\section{Continuing RHIC program} \label{continue}

In spring of 2008 a STAR spokesperson election led to new leadership and  dramatic rearrangements. The Event Structure working group was summarily disenfranchised, meaning that it was reduced to an ``interest group'' without the authority to advance research results to publication within STAR. In effect, that transition represented imposition of ``Bevalac rules'' across the collaboration. The novel physics results produced by the Event Structure working group over more than eight years were to be {\em overwritten} by the preferred flow/QGP narrative. As an example of what was at stake, one Event Structure paper (2006) was named ``paper of the year'' by the Journal of Physics G editorial board for revealing the complex relation between mean-pt fluctuations and jet angular correlations~\cite{ptscale}. That result then falsified the common assumption that mean-pt data variations might reveal collision system temperature trends. Such trends, as inferred from mean-pt data, were expected to support QGP claims~\cite{pttemp}.

A far more onerous consequence impacted juniors within the working group that had spent years pioneering novel analysis methods and  new insights across small and large collision systems. It soon became apparent that such innovative research could not emerge from the updated STAR collaboration as publications without a sort of ``poison pill'' approach. Text must be added to a paper to effectively neuter its main scientific conclusions. Struggles over that constraint continued long after affected juniors were forced to abandon the field. Some research results were published only in few-author papers external to STAR~\cite{v2ptb}.

\section{LHC commencement and surprises}

The experimental program of the CERN large hadron collider (LHC) commenced in 2010. Three large detectors support substantial heavy ion programs. ALICE was purpose built for the heavy ion program which is its dominant activity. ATLAS and CMS were purpose built for HEP research, most notably discovery of the Higgs boson (2012), but also include smaller heavy ion groups derived in part from within the RHIC collaborations.

An immediate and quite surprising CMS result was observation of a ``ridge'' in \pp\ angular correlations, a ``long range'' (on pseudorapidity $\eta$) structure  interpreted in \aa\ collisions to indicate collectivity (flow), but now appearing in the smallest collision system. The ``ridge'' was later observed in \ppb\ collisions as a double ridge, equivalent to the quadrupole structure\footnote{A quadrupole system may refer to a four-pole field ($+-+-$) whose azimuth $\phi$ distribution could be described by the function $\cos(2\phi)$ and thus correspond to the v2 measure.} represented by measure v2. Observation of collectivity in small collision systems appeared troublesome for a narrative in which collectivity (QGP) {\em requires} energetic central collisions of the largest nuclei, i.e.\ conditions only provided by energetic heavy-ion colliders.

The problem of ``QGP signals'' appearing inconveniently in small collision systems at the LHC has continued to fester over fifteen years. To be clear, since the late nineties the common intention  was that results from small systems representing ``cold nuclear matter'' were intended  to serve as a reference against ``hot QGP matter'' claims in \aa\ collisions. Substantial and consistent {\em differences} in various observables were supposed to support QGP claims. At RHIC that  seemed to be the case; results from d-Au (deuteron-gold) collisions seemed to offer marked differences and confirmation. However,  LHC found otherwise.

The response over time has been a manipulation of language that effectively abandoned the small-system reference or control experiment. For instance, there is now frequent referral to formation of ``QGP droplets'' (concept undefined) in systems as small as \pp\ collisions~\cite{phendroplets}. Confronted with the CMS ``ridge'' and other anomalous observations that have accumulated over years, a broadly cited review paper (2018) opines that ``...the field of relativistic heavy ion physics is in the midst of a revolution in our understanding of the conditions necessary for [QGP]. The revolution...[is] driven by  the experimental observation of flow-like features [i.e.\ ridges] in the collisions of small hadronic systems''~\cite{nagle}.

\section{The scientific method and formal logic}

Given such a major shift in scientific practice with the emergence of unexpected experimental results it is useful to review the concept of the scientific method. What is today recognized as the scientific method first emerged about four hundred years ago with the activities of Francis Bacon and Galileo Galilei (c.\ 1600). John Locke (c. 1660) advanced the concept of {\em empiricism} and encouraged anyone studying nature to {\em investigate for themselves rather than to accept the authority of others.} Knowledge should be based on empirical evidence, not {\em a priori} beliefs. Locke was closely associated with Robert Boyle and Robert Hooke (then Boyle's assistant) as they pioneered the practice of experimental science.

Within the modern method, scientific laws are based on reproducible experiments and openly available data. To qualify eventually as ``laws,'' theories or conjectures are tested by {\em independent scientists} whose results are published and therefore available to all for further debate. 
However, a third element of the method was only realized in the early twentieth century (c.\ 1930): {\em falsification} of a theory or conjecture. Attempting to prove a theory correct is misguided. Instead, if observations are inconsistent with a theory the theory must be discarded. 
A key element of the scientific method is then {\em skeptical challenge} wherein a scientific principle, {\em even if broadly accepted}, may be challenged at any time with contrasting evidence and possibly falsified by such evidence.

Returning to the LHC conundrum, the {\em scientific} response to small-system ``collectivity'' should have been to say either there is no QGP in {\em any} system (a simple response) or we don't understand the physics of nuclear collision data now available from the LHC (a productive response). The  review paper conclusion above says in effect ``magic happens'' as ``a revolution in our understanding.'' Extrapolated forward that principle suggests that we undergo a ``revolution in our understanding'' any time our checkbook doesn't balance (a dated reference).

To comment further on LHC data and logical structure, the ALICE collaboration has measured pt spectra for a variety of collision systems from \pp\ to \ppb\ to \pbpb. I have spent much of fifteen years studying the properties of \mbox{ALICE} data that are for the most part of good quality (the  published {\em data}). However, {\em interpretation} of spectrum data by ALICE within their publications is problematic. A major premise of the QGP ``signals'' approach is that radial flow (transverse expansion of the collision system due to pressure gradients via QGP) is manifested in spectra by ``hardening.'' The term refers to the rate of fall of spectra with increasing pt. ``Hardening'' means the rate of fall is reduced, presumably because the detected particles are emitted from a moving source.

However, spectrum ``hardening'' may also occur as a result of {\em jet production}. In fact, given what is now known about jet production in high energy nuclear collisions (a great deal) and how that has been understood in previous Event Structure spectrum studies, spectrum ``hardening'' is  more likely due to jet contributions to spectra. That is not a new idea. It traces back to 2006 and STAR publication of the Event Structure two-component model of \pp\ spectra~\cite{ppprd}.

ALICE spectrum papers comparing small and large systems include the following argument (paraphrased): ``We all know that QGP (i.e.\ radial flow) produces hardening in \aa\ spectra; we observe hardening in small-system spectra; therefore, there is QGP in small systems.'' Such arguments  contribute to the ``revolution in our understanding'' noted above. Stripping the claim down to bare logic ``A implies B, I observe B therefore A is true,'' where A is flow and B is hardening. That sequence represents a {\em formal fallacy}, where ``formal'' refers to the logical form of the statement. A valid alternative is ``A implies B, I observe {\em not}-B therefore {\em not}-A is true.'' This example---``Any Oreo cookie is round, this cookie is not round therefore it is not an Oreo cookie''--- is correct reasoning. This logical structure explains the importance of {\em falsification} noted above. The fallacious logical structure corresponds to attempts to prove a theory true. The valid example (regarding cookies) corresponds to falsifying a hypothesis. What {\em cannot} be so eliminated is (at least temporarily) accepted as valid.

Misunderstanding pt spectrum trends is just one example of faulty logic in play within the RHIC and LHC context. Another example relates to two-particle correlations.
``We show that anisotropies in transverse-momentum distributions [correlation structure on azimuth angle $\phi$] provide an {\em unambiguous signature} of transverse collective flow in ultrarelativistic nucleus-nucleus collisions [emphasis added]''~\cite{ollitrault}. ``Collective flow'' is then seen as a primary verification of QGP formation. In the first case A is ``collective flow'' and B is ``anisotropies.'' In turn, a related expression has A as ``QGP'' and B as ``collective flow'' thus revealing a {\em chain of fallacious argument}. In reality, several mechanisms {\em in combination} produce observed correlation structure, each with its own  systematic trends. To unravel the logical web the full suite of correlation {\em sources} must be identified  before claims of ``flows'' and ``QGP'' are addressed. That has been a target of Event Structure methods over ten years as described below.

\section{Continuing progress in general data analysis}

Event Structure methods have been applied to a number of major topics during LHC operation. What follows is a selected list of relevant papers: 

{\em Quadrupole spectra (2008)

Parton fragmentation to jets (2009)

Relevance of hydrodynamic theory to nuclear collisions (2010)

Relation of parton fragmentation to jet angular correlations (2011)

Comparing the CMS ``ridge'' to v2 measurements in Au-Au collisions (2011)

General description of jet production in nuclear collisions (2014)

Bayesian analysis of azimuth angular correlations (2015)

Relation of mean-pt fluctuations to jet production (2015)

Precision measurements of azimuth quadrupole in Au-Au collisions (2015)

Precision measurements of 200 GeV p-p 2D angular correlations  (2016)

Questioning QGP formation in small collision systems (2019)

Two-component model of pt spectra for p-Pb collisions (2020)

Two-component model of pt spectra for LHC p-p collisions (2021)}

\noindent The above texts paraphrase specific paper titles. Those topics confront the specific issues relating to QGP claims for small collision systems that converge below. Other papers addressing different topics were also produced. Major accomplishments include transformation of ``elliptic flow'' measure v2(pt) to quadrupole spectra on rapidity yt exhibiting great simplicity, clear understanding of pt spectrum evolution in all A-B collisions, accurate determination of jet contributions to spectra and full understanding of the CMS ``ridge'' based on Event Structure analysis of STAR 200 GeV p-p and Au-Au correlation data.

\section{Key developments and small/large  resolution}

The papers listed above partially summarize extensive efforts to establish a reliable context for analysis and interpretation of high-energy  collision data. Papers listed below then specifically target the central challenges presented by LHC heavy-ion-related data. Those papers are described in some detail.

\subsection{Collision geometry, PID spectra and jet quenching}

A fundamental issue for any collision system is determination of collision geometry: do two nuclei collide head on with almost all nucleons involved or only peripherally with interaction of a few nucleons each (collision centrality)? For a given geometric centrality how many nucleons interact and in what way?

{\em Glauber model geometry in p-Pb collisions at the LHC (2018)~\cite{tomglauber}

Discovery of ``exclusivity'' in nucleon-nucleon collisions (2018)}~\cite{tomexclude}

\noindent Those papers report that a Monte Carlo (random-number simulation) model of \ppb\ collisions (the ``conventional'' method) includes an unjustified assumption about nucleon-nucleon collisions {\em within} nucleon-nucleus (e.g.\ \ppb) collisions: that a nucleon may interact {\em simultaneously} with multiple other nucleons within the target nucleus. Empirically that is not valid. The result is overestimation of nucleon-nucleon collisions. The Glauber model may be improved by imposing an {\em exclusivity} constraint that nucleons interact with only one other nucleon {\em at a time}, which then introduces a new question: what clock measures that time?

{\em Accurate model for identified hadron spectra from p-Pb collisions (2021)~\cite{ppbpid}

Accurate model for identified hadron spectra from p-p collisions (2022)}~\cite{pppid}

\noindent Identified-hadron (PID) studies are required in part to resolve several longstanding issues having to do with ``hadrochemistry'' (e.g.\ what hadron species (types) are produced with what abundances in what A-B collisions?) Hadron species abundance is closely linked to QGP claims. Accurate species abundances also revealed new aspects of high energy collisions that remain under study.

{\em Nuclear modification factors for p-Pb collisions (2023)}~\cite{ppbnucmod}

\noindent ``Nuclear modification factors'' refers to ``jet quenching'' appearing in  text above. This is another case of logical fog. The  reference is to RAA, a ratio of AA spectra to pp spectra that is {\em rescaled} by a number Nbin supposed to represent the number of N-N collisions experienced by a nucleon passing through a target nucleus. For no jet quenching the ratio {\em at higher pt} should be near 1. However, that ``modification factor'' and its interpretations  plumb the depths of heavy-ion ignorance about jet production and collision geometry in high energy collisions. In particular, RAA conceals the presence of more than 95\% of jet fragments.

\subsection{Resolution of two major longstanding issues}

 The two papers below report a  resolution of some basic problems arising from newly established understanding of high-energy nuclear collisions. They respond to charges made within the STAR white paper to obtain clarification of the nature of elliptic flow and ``jet quenching'' and also to the problem of expected QGP signatures appearing in small collision systems. The first issue  is jet quenching and determination of the collision geometry of A-B collision systems.

{\em Space-time geometry of small and large collision systems at  LHC (2025)}~\cite{tompbpb}

\noindent That  study confronts the jet-quenching problem with new methods and last-minute insight. The TCM was applied to LHC \pbpb\ and \pp\ collisions. New methods for collision geometry inferred in 2018 led to  much-improved isolation of jet-related spectrum components. Key information emerged from observing that {\em before} any rescaling of spectra the \pbpb\ hard component {\em matches the \pp\ hard component} above a certain momentum (4 GeV/c). Reducing the \pbpb\ jet contribution to equal the measured \pp\ contribution exactly within data uncertainties via effects of a dense medium qualifies as a miracle. Resolution of the ``miracle'' emerged by reexamining the expression ``at a time'' noted above. 

Spectrum hard components arise from jet production resulting from parton interactions---large-angle scattering---within \nn\ interactions. The clock that determines when a nucleon might interact {\em again} may be governed by the speed of the interacting parton {\em within} a nucleon. Low-speed partons may produce low-pt hadrons whereas high-speed partons produce high-pt hadrons. The relevant ``clock'' of a parton depends on its speed relative to the lab frame. High-speed partons have slow clocks (relativistic time dilation) and may effectively ``freeze'' the parent nucleon long enough for it to pass through a target nucleus with only a single interaction.  That description is consistent with the new observations. ``Jet quenching'' within  a dense medium appears to be unnecessary. The second issue is appearance of a quadrupole contribution within small collision systems.

{\em Quadrupole spectra derived from Pb-Pb collisions (2025)}~\cite{newquadspec}

\noindent Demystification of the v2 system (hydro theory vs v2 data) has required a long and persistent effort. Major progress was made in 2008 with transformation of v2(pt) data to {\em quadrupole spectra}~\cite{quadspec}, the emission spectra of those hadrons exhibiting quadrupole angular correlations, based on certain relativistic transformations known since the seventies. A model spectrum shape in the correct coordinate system then described four hadron species, enabling prediction of v2(pt) data for any hadron species over a large pt range. That achievement also {excluded the  necessity} for any hydrodynamic theory to predict v2(pt) data.

The 2025 quadrupole paper~\cite{newquadspec} applies the same method to LHC \pbpb\ v2(pt) data with  similar results. In fact, the inferred 200 GeV \auau\ data trends also predict 2.76 TeV \pbpb\ data trends within their uncertainties. However, the major discovery within that study relates to {\em amplitude} variation of quadrupole correlations with collision system and collision geometry. A key element comes from a 2016 study of quadrupole amplitudes in RHIC \pp\ collisions~\cite{ppquad}. Amplitude measured by {\em number of correlated hadron pairs} increases exactly as the {\em cube} of the multiplicity soft component, which in turn is understood as the number of low-speed  partons, essentially all gluons. Thus, at least in \pp\ (small) collisions the quadrupole is generated by {\em three-gluon interactions}.

The final step required in effect a dive off a high platform. Within \pp\ collisions the Event Structure group had observed a fixed relation between the TCM soft component (varying linearly with gluon number) and the hard component (varying as the square of the gluon number, consistent with dijets  from two-gluon interactions). The {\em product} would then result in a cubic trend as in Ref.~\cite{ppquad}. Could that relation be applied to {\em any} A-B collision system? The implementation required accurate estimation of collision geometry: the number of nucleon participants and their binary collisions. The several results above were combined to predict a trend for the quadrupole amplitude in {\em any} \mbox{A-B} collision system. The combination accurately predicted quadrupole  variation over {\em six orders of magnitude} from RHIC \pp\ to LHC central \pbpb, including RHIC \auau\ data. A simple production mechanism was quantitatively related to v2(pt) measurements across  A-B collision systems and a large energy interval.

\section{A simple system -- QCD in the real world}

A more-recent paper discussed below presents a  summary of the current understanding of high-energy nuclear collisions arising from Event Structure methods applied to RHIC and LHC data.

{\em Few-gluon interactions and multipole radiation in high-energy collisions}~\cite{jetspid}

\noindent The paper first introduces three review articles widely considered to summarize a {\em group consensus} regarding QGP discovery. The second part of the paper emphasizes the copious evidenced for strong jet contributions to spectra, correlations and yields down to low pt and the pitfalls of analyzing and interpreting those measures within a flow/QGP context where jets are assumed to play a minor role, and only at high pt. ``Multipole'' refers to dipole, quadrupole, etc.

The third part of the paper considers two-particle correlations and especially the recent results from Ref.~\cite{newquadspec} on v2 data reinterpreted as a {\em color quadrupole} manifestation, including the  prediction of a quadrupole amplitude trend applied to all collision systems in common. The novel CMS ``ridge'' is then subsumed within a simple Event Structure paradigm.
Reference~\cite{jetspid} applies similar analysis to {\em color dipole} (dijet) amplitudes, also describing all collision systems, with a predicted trend for jet production extending over four orders of magnitude.

Reference~\cite{jetspid} combined with Ref.~\cite{tompbpb} provides an alternative description of hard-component (jet) contributions to spectra {\em excluding} ``jet quenching.'' It reports recent developments that provide a  simple picture of high-energy collisions at the RHIC and LHC. Collision dynamics lead to hadron production via three elementary QCD mechanisms: (a) projectile-nucleon dissociation varying proportional to  the number of slower or low-$x$ gluons near midrapidity, (b) two-gluon interactions leading to color dipole  radiation (dijets) proportional to gluons squared and (c) three-gluon interactions leading to color quadrupole radiation varying as gluons cubed. The typical participant nucleon densities estimated from {\em measured} spectrum properties (mean pt) are much lower than what has been estimated from conventional numerical simulations, inconsistent with the ``hot and dense'' scenario usually associated with QGP formation.

The last part of Ref.~\cite{jetspid} returns to the three review articles in the first part. It contrasts specific claims quoted there to support QGP discovery with recent results via Event Structure methods  summarized above. Dramatic contradictions between  two approaches are apparent. Event Structure methods have been applied since the first days of RHIC operation. Analysis results emerging from two systems  then  diverged monotonically with time over twenty-five years. Recent Event Structure results summarized here have resolved a growing crisis.

\section{Cult science versus the scientific method}

This section considers a larger philosophical context: how social and psychological elements affect scientific validity. A comparison is made between a hypothetical ``cult science'' scenario as described by Richard Feynman and a scenario more consistent with the scientific method as it is at least nominally understood.

\subsection{Large collaborations and cult science}

``Cult science'' refers here not to a religious cult but to the product of a community that believes it is conducting legitimate scientific research but fails due to systematic imperfections in its approach. The term ``cult'' used here relates to the expression ``cargo cult science'' introduced by Richard Feynman in 1974~\cite{cargocult}. 

The approach to research adopted by the heavy ion community during RHIC and later LHC operations seems to have the requisite features of a major research program: massive state-of-art detectors,  acquisition of large volumes of particle data, application of a variety of statistical analysis methods,  production of complex and varied data representations, preparation of many papers for publication and presentation of results at major scientific conferences.

However, as often happens the devil is in the details. In contrast to John Locke's admonition that scientific advances should be {\em based on individual observations not subject to the authority of others}, the structure of a large accelerator collaboration (hundreds of members) is inherently pyramidal: a small number of seniors is in charge while a large number of persons (many of them students) must submit to authority, whatever the system's {\em nominal} structure description.

The complex collaboration {\em infrastructure} described above presents the {\em facade} of a scientific process that produces a large volume of raw particle data. The subsequent processing however has been deeply flawed.
Technical language and analysis methods continuing up to the present tend by convention to bias toward a preferred outcome. The term ``collectivity'' is assigned to almost any correlation structure, reflecting the Bevalac experience. The term ``ridge'' is applied to correlation structure extending over some distance on angle measure $\eta$ and thus interpreted as evidence for ``collectivity.'' In turn, ``collectivity'' is translated (again per Bevalac experience) to ``flow'' taken as evidence for QGP.

Preferred analysis methods tend to {\em minimize} the information carried by data, as in spectrum ratios that confuse distinct collision mechanisms, or present overly-complex schemes that cannot be evaluated by ``outsiders'' as a means of skeptical challenge. For instance, the many v2 ``methods'' (why not just one?)  may include varying jet (``nonflow'') contributions as unknown systematic errors.
Analysis methods that ``get the right answer'' are approved and {\em protected} from skeptical challenge. Methods that appear to ``get the wrong answer'' (i.e.\ conflict with the preferred narrative) are excluded from further engagement even if fully competent. Over time such trends amount to a ratchet process that worsens bias and eliminates the possibility for effective research. What emerges serves as a psychological projection screen for individual and group self deception.

The result is just the situation that Feynman warned against in 1974: Instead of searching carefully for any possible evidence to {\em disprove the preferred narrative} per the scientific method, the preferred narrative, what might be compared to a cult ideology, is consistently and {\em stridently} protected including imposing strong procedural constraints on collaboration activity as in Sec.~\ref{continue}.

Such trends extend beyond collaboration boundaries. Scientists working alone or in small groups encounter the challenge of publishing their work in major journals if their results run counter to the preferred narrative. Publication requires peer review by referees intended to ensure that what appears in a journal is competent science. But, in the context relevant to this article, referee candidates may be drawn from a population corresponding either to collaboration membership or to a theory community that may have {\em generated prominent elements of the narrative itself}. Ideally, referees should adhere to the standards advertised by a journal and evaluate a submitted manuscript fairly, competently and without regard to whether they {\em agree with its conclusions or not.}

The last point is of central importance. Scientific {\em debate} implies that the debaters {\em do not agree} on significant elements of science. Debate is intended to resolve differences by reducing conceptual confusion or suggesting new experimental approaches to resolve ambiguity. But if a referee's primary criterion for paper acceptance is agreement with his own beliefs then the system is compromised regarding the scientific method. Essential skeptical challenge is effectively blocked. Evidence for that reality may be found in the bibliography of this article where a substantial number of papers remain unpublished after years.

\subsection{Individual scientists and the scientific method}

Locke's admonition that the scientific method should be based on the individual scientist and his observations is the basic element of the process. The general intent is to understand nature as well as possible given available resources. To paraphrase Feynman, a scientist makes some observations then {\em guesses} what their meaning is. The guesses (conjectures, theories) are then subjected to {\em skeptical challenge}, tested for correctness against further observations. Inevitable errors must be corrected. The cycle of conjecture, experiment and challenge moves  understanding (science) forward in an evolutionary process.

For that process to work the scientist must be scrupulously honest, at least with regard to science. Yes, the guess was wrong this time so start over. Perhaps the observations were flawed so repeat them, but that cannot excuse an erroneous conjecture. Errors occurring at a certain rate are inherent in the process. A scientist should be able to ``grade his own homework,'' that is, rigorously examine his own product for errors and {\em demonstrate skill in finding them.}

The Event Structure working group was blessed from the beginning with junior members who brought sophisticated new skills to nuclear physics. In the late nineties personal computers (PCs) within private offices first became available (no more carrying a box of Hollerith cards to the computer center). Sophisticated new computer languages (ANSI C and C$^{++}$) became the mainstream. Analysis of RHIC data faced major  challenges: processing {\em terabytes} of particle data on processor farms consisting of hundreds of computers, tracking the computation output directed to  higher-level analysis by a data base (in effect, catalog system) that had to be invented. Junior members introduced a variety of methods on the fly to {\em ensure the integrity of the analysis system.}

At the same time, junior members undertook to develop methods required to make sense of RHIC data {\em without a priori assumptions}, again following Locke and disregarding community prejudice. All aspects of data analysis were reconsidered and improved so as to {\em maximize} information extraction from data and present it in the most interpretable formats. In deference to Feynman, self-skepticism was encouraged---a routine question was what mistakes did we make today---along with the skills to fix mistakes and continue with progress.
Ideally, that process for individuals would then proceed to a broader level with debate among colleagues. The debate process would require {\em collegiality} to be effective. Skeptical challenge must be {\em welcomed} as producing better science for all.

However, as alluded to in Sec.~\ref{continue} the broader STAR collaboration grew increasingly hostile to Even Structure methods as their results diverged from the preferred narrative. Without {\em ad hoc} limits imposed on data ranges, abundant low-energy jets became apparent in contradiction to thermalization in a dense medium. v2 data plotted within more-transparent formats did not conform to expectations for a flowing medium. Increasing resistance emerged from management to publication of Event Structure results or presentation at conferences.

In contrast, results seeming to support the preferred narrative were not only published promptly, with corresponding talks positioned prominently at international conferences, but also featured within media pieces on the web pages of collaborations and national laboratories. There is thus a  {\em propaganda} advantage held by ``cult science'' over individual scientists practicing the scientific method. Such trends relate to another comment by Feynman: ``It's a kind of scientific integrity... if you're doing an experiment, you should report everything that you think might make it invalid...other causes that could possibly explain your results.'' That would imply featuring {\em unfavorable} evidence as prominently as {\em favorable} evidence. Feynman's 1974 address to Caltech students cited above ends thus:
``I have just one wish for you---the good luck to be somewhere where you are free to maintain the kind of integrity I have described, and where you do not feel forced by a need to maintain your position in the organization, or financial support, or so on, to lose your integrity. May you have that freedom.''
50 years later it seems that subsequent events have not favored Feynman's admonitions.

\begin{appendix}

\section{Thomas A. Trainor}

Thomas A. Trainor is  research professor of physics emeritus at the University of Washington. He is currently credited as  author or coauthor of over 350 papers addressing various topics in nuclear and atomic physics and the study of fundamental symmetries. Since RHIC startup in 2000 he has produced almost 100 single- or few-author papers addressing statistical analysis and interpretation of high-energy nuclear collision data from  RHIC and LHC collaborations.

Trainor has organized or co-organized a number of international conferences on statistical analysis and interpretation of data from high-energy nuclear collisions. He was convener of the Event-by-Event working group and then co-convener of the Event Structure working group within the STAR collaboration at the RHIC. He was also a member of the STAR Council governing body.

Since 2009 Trainor has been a member of the Board of Elders (steering committee) of the International Symposium on Multiparticle Dynamics (ISMD) with its fifty-plus year history of the study of high-energy nuclear collisions.

\section{Technical details}

This document is intended for non-specialists. Following are  technical details. The structure of high-energy nuclear collisions and resulting particle distributions is described in relation to the A-B collision (beam) axis. Particle momentum perpendicular to the axis is represented by  transverse momentum with symbol pt. Azimuth angle about the beam axis is represented by symbol $\phi$, and  polar angle ($\theta$) relative to the axis by related quantity $\eta$ (pseudorapidity). For a collider accelerator (where two beams travel in opposite directions and intersect at specific crossings), detector systems define the laboratory (lab) frame.

A-B collision geometry (e.g.\ \aa\ centrality) is described in terms of a number of participant nucleons Npart and a number of nucleon-nucleon (N-N) binary collisions Nbin related to some event-wise observable quantity, for instance the total particle count within some fraction of the total angular space. Geometry determination has been estimated conventionally by numerical simulations (so-called Glauber model). However, geometry estimated directly via measured quantities (e.g.\ mean pt) appears to provide more-accurate information.

Within a high-energy nuclear collision nucleons may be represented in terms of their partonic structure (partons are quarks and gluons). Particle production is determined in part by nucleon longitudinal structure (distribution of partons) on nucleon momentum fraction x (i.e.\ the fraction of nucleon total longitudinal momentum carried by an individual parton). The parton distribution on x is determine by collision energy and by a {\em splitting cascade} wherein partons split to form daughter partons (quark splits to quark plus gluon, gluon splits to gluon plus gluon) to form a tree-like structure. The depth of the cascade depends on collision energy. Particles (hadrons) produced at midrapidity in the lab frame emerge from the bottom of the cascade, {\em mainly from low-x gluons}.

Particle spectra are plotted in their basic form as frequency (number of particles per collision event) vs pt. Other choices for independent variable besides pt include transverse mass-energy mt and transverse rapidity yt. There are several mathematical models for particle spectra, with strongly debated interpretations. Particle spectra tend to approximate an exponential decrease at lower pt transitioning to a ``power-law'' trend at higher pt. A purely thermal system would follow an exponential trend very closely (e.g.\ the solar spectrum). Ensemble-mean pt is an important spectrum parameter relating to two or more hadron production mechanisms (e.g.\ leading to soft and hard components).

Data analysis emphasizes determination of the relation between observed phenomena (specific features manifested by data) and collision geometry or energy. For instance, how does jet production increase with event multiplicity?

Some papers listed below may be accessed in PDF form using the link
 
https://inspirehep.net/literature?sort=mostrecent\&size=25\&page=1\&q=

\noindent by inserting the included arXiv number, e.g.\ arXiv:2510.05314, in the entry slot.

\end{appendix}

\end{document}